\documentclass[usenatbib,useAMS]{mn2e}
\pdfoutput=1

\usepackage{graphicx}

\usepackage{natbib}
\usepackage{verbatim}

\newcommand{\sub}[1]{\ensuremath{_{\mbox{#1}}}}

\newcommand{\EE}[1]{\ensuremath{\times 10^{#1}}}
\newcommand{\degrees}{\ensuremath{^\circ}}
\newcommand{\keV}{\ensuremath{\mbox{ keV}}}

\newcommand{\erg}{\ensuremath{\mbox{ erg}}}
\newcommand{\s}{\ensuremath{\mbox{ s}}}
\newcommand{\Hz}{\ensuremath{\mbox{ Hz}}}

\newcommand\araa{{ARA\&A}}%
\newcommand\apj{{ApJ}}%
\newcommand\apjl{{ApJ}}%
\newcommand\apss{{Ap\&SS}}%
\newcommand\aap{{A\&A}}%
\newcommand\mnras{{MNRAS}}%
\newcommand\nat{{Nature}}%
\newcommand\gca{{Geochim.~Cosmochim.~Acta}}%
\title[Ultraviolet Spectra of GRBs from Massive Rapidly-Rotating
Stellar Progenitors]{The rest-frame ultraviolet spectra of GRBs from massive rapidly-rotating stellar progenitors}

\author[Peter B. Robinson, Rosalba Perna, Davide Lazzati, and Allard J. van
Marle]{Peter B. Robinson$^{1}$, Rosalba Perna $^{1}$, Davide Lazzati$^{2}$, and
Allard J. van Marle$^{3,4}$. \\ $^{1}$JILA,University of Colorado, 440
UCB, Boulder, CO 80309-0440\\ $^{2}$Department of Physics, NC State
University, Campus Box 8202, Raleigh, NC 27695-8202\\ $^{3}$Bartol
Research Institute, University of Delaware, 102 Sharp Laboratory,
Newark, 19716 DE, USA\\ $^{4}$Presently residing at: Centre for Plasma
Astrophysics, K.U. Leuven (Leuven Mathematical Modeling and
Computational Science Center),\\ Celestijnenlaan 200B, 3001 Heverlee,
Belgium}

\begin{document} 
\date{} \pagerange{} \maketitle

\begin{abstract}

The properties of a massive star prior to its final explosion are
imprinted in the circumstellar medium (CSM) created by its wind and
termination shock.  We perform a detailed, comprehensive calculation
of the time-variable and angle-dependent transmission spectra of 
an average-luminosity Gamma-Ray Burst (GRB) which explodes in the CSM
structure produced by the collapse of a 20 $M_{\sun}$, rapidly
rotating, $Z=0.001 $ progenitor star.  We study both the case in which
metals are initially in the gaseous phase, as well as the situation in
which they are heavily depleted into dust.  We find that high-velocity
lines from low-ionization states of silicon, carbon, and iron are 
initially present in the spectrum only if the metals are heavily
depleted into dust prior to the GRB explosion. However, such lines
disappear on timescales of a fraction of a second for a burst observed on-axis, 
and of a few seconds for a burst seen at high-latitude, making their
observation virtually impossible. Rest-frame lines produced in the
termination shock are instead clearly visible in all conditions. 
We conclude that time-resolved, early-time spectroscopy is not a
promising way in which the properties of the GRB progenitor wind can be
routinely studied. Previous detections of high velocity features in
GRB UV spectra must have been due either due to a superposition of a physically
unrelated absorber or to a progenitor star with very unusual properties. 
\end{abstract}

\section{Introduction}

For a few hours after their onset, the afterglows of Gamma-Ray Bursts
(GRBs) are the brightest sources in the far Universe. Their high
luminosity, together with their powerlaw, featureless spectrum, make
them ideal sources to probe their surrounding environment through the
absorption lines imprinted in their spectra.

Theoretical studies have suggested that long Gamma Ray Bursts
(GRBs) are produced by the collapse of rapidly-rotating, chemically
homogeneous, massive stars (e.g. Macfadyen \& Woosley 1999). The
association between GRBs and massive stars, which has been
observationally supported \citep{Stanek_etal03, Hjorth_etal03}, makes
absorption studies potentially useful as a new way to probe
star-forming regions at intermediate and high redshifts, and/or the
last hundreds of years of the progenitor evolution. Absorption lines
imprinted by the material ejected by the star prior to its explosion
allow one to probe the velocity structure and metal content of the
ejecta.  High-resolution spectroscopy of GRB afterglows has been
performed in a number of cases (e.g. \citet{Moller_etal02,
Matheson_etal03, Mirabal_etal03, Schaefer_etal03, Starling_etal05,
Fiore_etal05, D'Elia_etal07, Chen_etal08}; Prochanska et
al. (2008a), (2008b); Fox et al.  (2008); Thoene et al. (2008); see
also Whalen et al. 2008 for an extended discussion on absorbers in
GRBs), yielding constraints on the nature of the absorbing medium.
An especially well studied burst was GRB~021004, whose high-resolution
spectroscopy revealed a complex velocity structure of the absorbing
material, with velocities up to $\ga 3000$ km/s. The interpretation of
these features has been controversial.  Initial studies claimed
that the high velocity lines were a direct proof of the association of
GRBs with WR stars. \citet{Starling_etal05} argued that the lines
must be produced in a fossil stellar wind with hydrogen enrichment
from a companion. \citet{Mirabal_etal03} and \citet{Schaefer_etal03}
interpreted the lines as the result of shells of material which are
present around the progenitor.

More recent studies have however cast doubts on the initial
interpretation. \citet{Lazzati_etal06} performed a detailed time
dependent analysis, taking into account the burst flash
ionization. Even though they still considered the wind of the WR
progenitor star as the best absorber candidate, they pointed out that
such an interpretation would require a termination shock at a distance
of at least 100~pc. Such a large radius of the termination shock is
somewhat at odds with current wind models and could be accounted for
only if the progenitor were an extremely massive star evolving in a
fairly low-density environment.  \citet{Chen_etal07} studied a sample
of 5 GRBs with high velocity features (among which GRB~021004). Using
fine-structure transitions and the presence of low-ionization species,
they argued that the location of the absorber is very likely at a
large distance from the burster ($>1$~kpc), favoring an intervening
halo, physically unrelated to the GRB, as the location of the
absorbing material. Similarly, \citet{Vreeswijk_etal07} modeled fine
structure lines present in the spectrum of GRB 060418 with a UV
pumping model and concluded that the absorbers are at a distance of
1.7 $\pm$ 2 kpc. These results are consistent with a GRB completely
ionizing the local CSM, and with high-velocity absorbers elsewhere in
the host galaxy or intervening intergalactic space (see also
Prochanska et al. (2008a,b) and D'Elia et al. (2009a) for similar
findings in the case of other GRBs they studied).

Most of these studies do not take into account the possible shielding
effects of an unusually dense CSM,  as well as any effects due to dust.
Additionally, when photoionization calculations are performed, most
assume a steady state solution, whereas the radiation fields of GRBs are
highly temporally variable, with initial burst timescales of a few to
tens of seconds, and the afterglow varying over minutes to hours.  
Disentangling the physical origin of the spectral absorption features
seen in GRB afterglows is clearly of great importance for a full
understanding of the GRB phenomenon.

The aim of this paper is a detailed, systematic study of the
time-dependent transmittance spectrum of a GRB exploding in a 
dusty circumburst medium (CSM) shaped by its progenitor star. Recent
studies have suggested that long GRBs are produced by the collapse of 
rapidly-rotating, chemically homogeneous, massive stars
\citep{YoonLanger05, WoosleyHeger06, Yoon_etal06, Cantiello_etal07}.
This kind of evolution is believed to take place at low metallicity,
where stellar winds are weak and angular momentum loss by winds is
marginal.  The wind of numerous types of GRB and SN progenitors has been
simulated by \citet{vanMarle_etal08}, including one model which
recreates an anisotropic wind close to the star. This theoretical
scenario has found support in recent observations.
\citet{Campana_etal08} performed a detailed study of the X-ray spectrum
of GRB~060218, discovering a larger than normal N/O ratio in the
surrounding of the burst progenitor. They concluded that only a
progenitor star characterized by a fast stellar rotation and sub-solar
initial metallicity could produce such a metal enrichment.

In this paper we perform a detailed and comprehensive study of the
effects of the GRB and afterglow radiation on the CSM structure produced
by the stellar wind of a rapidly-rotating, low-metallicity, massive
progenitor star \citep{vanMarle_etal08}.  We aim at predicting the
time-variable, angle-of-sight dependent transmittance spectra of a
GRB afterglow which resulted from the collapse of such a star. We study
two cases, that of a CSM whose metals are initially in the gaseous phase
(i.e. no dust), and that of a CSM whose metals are heavily depleted into
dust prior to the GRB explosion.  Even though the formation and
survival of dust in the wind of Wolf-Rayet stars may seem surprising,
dust is observed in WR winds \citep{Allen_etal72,Williams_etal90}. The
gas temperature and the exposure to the star's UV radiation make the
wind a very harsh environment for dust particles. If the dust were to
share the temperature of the gas phase, the grains would quickly
sublimate. One key ingredient that allows the dust to survive is its
capability of radiatively cooling to temperatures lower than the gas in
which it is embedded (e.g. \citep{Lazzati08}). Even so, there are still
many unknown riddles in the way in which dust particles are born and
survive in WR winds, but their presence has been clearly detected (e.g.
Crowther 2003).

Although it is presently unknown whether the particular stellar
evolution model that we consider here is truly representative of
critically rotating stars in general, we believe that the
ability to compare theoretical absorption features with observational
data can be potentially useful to the study of both GRBs and
massive star evolution.

\section{The CSM of the Progenitor Star and  Spectra Calculations}
\label{Methods} \subsection{The CSM of the Progenitor Star}

Van Marle et al. (2008) computed the evolution of the circumstellar
density profile around a rapidly rotating, $20 M_{\sun}$, low
metallicity star ($Z=0.001 $) star, which is considered representative
of a typical GRB progenitor. Due to the anisotropy associated with the
rotation of the star, the calculation yielded density profiles which are
dependent on the angle $\theta$ between the rotation axis (presumably
coinciding with the GRB jet axis), and the line of sight.

We choose 3 different radial profiles to show the angular and radial
dependence in detail, at $\theta = 0$, $25$, and $80$ degrees. The
calculations of \citet{vanMarle_etal08} provide the density, the radial
and tangential velocity $v_r$ and $v_\theta$, the ionization state of
hydrogen, and the internal energy as a function of radius $r$ and angle
$\theta$. The internal energy is used to compute the initial temperature
as a function of radius by means of the ideal gas equation of state with
the polytropic index $\gamma = 5/3$. The initial ionization state of
elements heavier than hydrogen is computed using the photoionization
code of \citet{PernaLazzati02}, assuming photoionization equilibrium
under the blackbody radiation field from a star with a characteristic
temperature of $1.4\EE{5}$ K and a luminosity of $4\times 10^6
L_{\sun}$, which correspond to the properties of the star at the end of
its life as given by \citet{vanMarle_etal08}.

The density profile is assumed constant in time throughout the
calculation, which is reasonable since the hydrodynamical timescales are
much longer than the ionization timescales.  For the same reason, the
radial velocities of the gas are assumed constant in time as well.

For the metal content of the CSM, we use the metallicity history of the
stellar photosphere as calculated by \citet{Yoon_etal06}. This, combined
with the radial velocity information, is used to compute a
radially-dependent metallicity profile.  The outer shell, with near-zero
radial velocity, has the initial metallicity (75\% H 25\% He with trace
elements), while the inner region, dominated by the stellar wind, is
very helium enriched. The helium enrichment is due to the rapid
rotation of the star; it causes the star to evolve homogeneously because
of mixing, so that at the end of the star's life nearly all of the
hydrogen has burned into helium. 

For the species that are not included in the calculations of
\citet{Yoon_etal06} (Sulfur, Argon, Calcium, and Nickel), we assumed a
metallicity of $Z=0.001$ with the relative weights of the species as in
the solar case \citep{AndersGrevesse89}, for both regions of the CSM.
The abundance values we used for all 12 species are given in Table
\ref{Metallicities}. The HI column density is about $1.5\times 10^{21}$ 
cm$^{-2}$.

\begin{table} 
\caption{Abundances used for the outer shell and
the inner region. These are given on a base 10 logarithmic scale
normalized so that the abundance of hydrogen is 12.  } 
\begin{center}
\centering 
\begin{tabular}{cccc} \hline\hline 
Element &  Outer Shell Abundances & Inner Region Abundances \\ 

He & 10.90 & 13.84\\ C  & 7.28 & 8.64 \\ N  & 6.68 & 10.33 \\ O  & 7.61
& 8.62 \\ Ne  & 6.75 & 8.92 \\ Mg  & 6.17 & 8.47 \\ Si   & 6.20 & 8.52
\\ S   & 5.9 & 5.9 \\ Ar   & 5.6 &  5.6 \\ Ca  & 5.0 & 5.0\\ Fe  & 6.19
& 8.51 \\ Ni  & 5.0 & 5.0 \\ 
\end{tabular} 
\end{center}
\label{Metallicities} 
\end{table}

\begin{figure} 
\begin{center} %
\includegraphics[scale =.8]{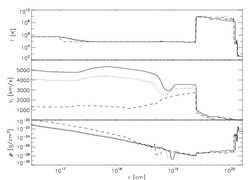} 
\caption{Temperature, Radial Velocity,
and Density Profiles.  The solid, dotted, and dashed lines represent
the $\theta = 0,25,$ and 80$^\circ$ lines of sight, respectively. }
\label{InitProfile} 
\end{center} 
\end{figure}

\begin{figure} 
\begin{center} 
\includegraphics[scale =.95,trim=3cm 0 0 0, clip] {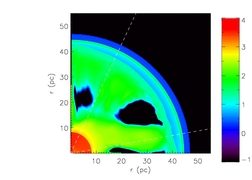} 
\caption{Radial Velocity Map: The radial velocity map for the polar 
region between
angles $\theta=0$ and $\theta=\pi/2$ rad. The color scale is in units of
log [km/s].  The 0-10 pc region has velocities ranging between 2730 and
5332 km/s, with the highest speeds found at smaller polar angles.
The over-plotted white dash lines indicate the lines of sight displayed
in Fig.\ref{InitProfile}. Note that one line of sight lies along the
pole (y axis) and that positive velocity here indicates outflow. }
\label{Velocity2D} 
\end{center} 
\end{figure}

\begin{figure} 
\begin{center} 
\includegraphics[scale =.95,trim=3cm 0 0 0, clip]{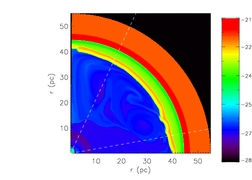} 
\caption{Density Map:
Similar to Fig. \ref{Velocity2D} but for the density of the medium. The
color scale is in units of log [g/cm$^{3}$]. } 
\label{Density2D}
\end{center} 
\end{figure}

\begin{figure} 
\begin{center} 
\includegraphics[scale =.95,trim=3cm 0 0 0, clip]{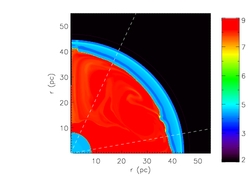} 
\caption{Temperature
Map: Same as in Figs. \ref{Velocity2D} and \ref{Density2D}, but for the
temperature of the CSM prior to the star explosion. The color scale is
in units of log [K]. } 
\label{Temperature2D} 
\end{center} 
\end{figure}

The velocity, density and temperature profiles produced by the
simulations above are displayed in Fig.\ref{InitProfile} for the three
lines of sight under consideration.  Full 2D plots of the velocity,
density, and temperature are shown for completeness in
Fig.\ref{Velocity2D}, \ref{Density2D}, and \ref{Temperature2D}. These
figures display the physical conditions of the CSM just as the star is
about to explode.

An component which is not included in these models is dust. As a
matter of fact, while it is known that dust is produced by stars,
there is very little information on the precise radial distribution,
grain-size distribution and composition of the dust grains in the
close CSM of the star. Therefore, in our study we consider two
limiting situations, one in which there is no dust (all the metals are
in the gaseous phase to start with), and one in which the metals are
heavily depleted into dust at all radii. In this case, we take a dust
population similar to that of the Milky Way, consisting of graphite
and silicate grains with a power-law distribution in size with slope
-3.5, and a range of sizes between .005 $\mu$m and .25 $\mu$m. Since
the metallicity of the CSM is radially dependent, we determine at each
radius the dust-to-gas ratio by assuming that 90\% of the least
abundant element among the ones making up the grains (C, Mg, O, Si,
Fe) is depleted into dust.  We point out that the precise details of
the initial dust distribution are not very important for our results,
since the dust grains close to the GRB location are quickly destroyed
by the intense radiation from the burst. Similarly, any
uncertainty in the destruction rates by a factor of a few would not
affect sensibly our results, since recycling is extremely fast with
respect to photoionization.  However, there is a fundamental
difference between an initial condition with no dust and an initial
condition with dust. In the former, low-ionization species such as
CII, FeII, SiII are photoionized by the strong radiation field of the
progenitor star in its close CSM, where the high-velocity,
free-streaming wind resides. Therefore, high-velocity lines from these
elements are never visible. In the latter case, these elements are
initially shielded within the dust grains. Then, when the star
explodes and the prompt GRB and afterglow radiation impinge on the
grains, those depleted elements are recycled into the medium, and they
are visible for a period of time until they get photoionized. Thanks
to the use of the code by Perna \& Lazzati (2002), we are able to
model self-consistently the destruction of the dust grains, their
redistribution into the CSM, and their subsequent
photoionization. When there is dust, high-velocity lines from
low-ionization species become visible\footnote{We need to point
out that the results we display in the following would change
qualitatively if the dust composition were radically different
from the MW distribution assumed here. For example, if the grains,
instead of being composed of Fe, Mg, Si, O and C  (as in
the MW dust distribution), were made up of other elements, then lines
from those elements would be seen coming from the wind.}

We thus have all the initial conditions of the medium at the time at
which the star explodes and produces a GRB: temperature, density, radial
velocity, metallicity and dust as a function of radius and polar angle.
These conditions provide the starting point for the computation of the
time-dependent transmittance spectra expected during the GRB afterglow
phase.

\subsection{Numerical Setup and Radiative Transfer}

We divided each line of sight (the three chosen values of $\theta$) into
1909 radial bins, matching the resolution of the data generated by
\citet{vanMarle_etal08}. In the time coordinate, we chose
logarithmically spaced time steps starting at $10^{-5}$ seconds and
ending at a time of 3 days. It's worth noting that time $t = 0$ is
defined as when the first photons reach the observer. Thus, observational
effects like superluminal velocities of ionization fronts are apparent,
not actual.  Ionization profiles are, instead, in the rest frame of the
burst.  Complete ionization profiles (necessary for generating the
spectra) were therefore extracted as a function of time.

The simulation of time-dependent photoionization and radiation
transfer was performed using the code by \citet{PernaLazzati02} (see
also \cite{PLF03}; \cite{LazzatiPerna03}).  This code keeps track of
the 13 most abundant elements (H, He, C, N, O, Ne, Mg, Si, S, Ar, Ca,
Fe, Ni) and computes their ionization state as a function of time.
The continuum opacity is also part of the computation, while the
opacity due to resonant lines is calculated separately (see \S
\ref{transcomputation}). The radiative transfer is performed in the
$1\,{\rm eV}-10 \,{\rm keV}$ spectral range.

We note that, once the ions from the sublimated grains return to
the gaseous phase, they are given the temperature of the gas at that
location. This is likely to be inaccurate, but unfortunately, at this
stage it is not possible to implement a more accurate prescription in
our code for both a technical and a theoretical limitation.
Technically, once the ions are ejected from the dust, they are not
distinguishable (in the code) from the gas ions. The code does not
allow for two components of the same ion at different
temperatures. Furthermore, even if the code were to include this
possibility, there is not a single dust temperature. The temperature
of the sublimating dust grains depends on the grain size and on the
distance to the burst. Therefore one would need to consider a
continuum of temperatures and the problem would become numerically
untreatable. Finally, there is a theoretical uncertainty in what is
the distribution of the ejected ion energies. One possibility is that
the ions are ejected with a thermal energy distribution corresponding
to the temperature of the dust. The issue with this procedure is that
the ejected particles have to overcome a potential barrier (the
binding energy to the grain) to leave the dust grain, and there is no
clear recipe on how to compute the ejected particle energy
distribution. For big grains, it is likely to be an exponential tail
(and therefore non-thermal), but for small grains quantum effects are
important. We consider a detailed implementation of all these effects, albeit
interesting, beyond the scope of this paper.

\subsection{The Light Curve of the GRB and its Afterglow}
\label{lumfunctionsection}

The light curve of the GRB and its afterglow was modeled with the sum of
two separate components: a prompt emission component and an afterglow
component. The models we adopted are described in this section. For
each, we let our light curve take the form: 
\begin{equation}
L_{\nu}\left(t,\nu, \theta \right) = A T(t)
F_\nu(\nu)\left(\frac{\theta_0}{\theta} \right)^2\;,
\label{GeneralLightCurve} 
\end{equation} 
where $A$ is an appropriate normalization constant, $T(t)$ contains the
time dependence of the radiation, $F_{\nu}(\nu)$ is the flux at a given
frequency $\nu$, and $\theta_0$ represents the half opening angle of the
jet core within the context of a structured jet
\citep{Rossi_etal02,ZhangMeszaros02_2}. This is observed to vary from
less than 1$\degrees$ to as much as 15$\degrees$
(Panaitescu 2007; Liang et al. 2008) and is taken here to be
$\theta_0=2.5^\circ$ . Note that the angular dependence of the
luminosity does play a significant role in the ionization timescale of
the various elements at different polar angles.

Both the spectral function $F_\nu(\nu)$ and the time function $T(t)$ are
normalized so that the integrals over frequency and time, respectively,
are equal to 1, making the normalization of the luminosity entirely
contained within the constant $A$.  Additionally, we note that
\begin{equation} 
\int_0^\infty L_\nu \left(t , \nu ,\theta \right)  d\nu
= L_{\rm iso}(\theta) \label{LCintdnu}\;, 
\end{equation} 
where $L_{\rm iso}$ is the effective isotropic luminosity for the line
of sight we are interested in, an important input onto the radiative
transfer code.

\subsubsection{ The Prompt Emission} For the GRB prompt emission, we
choose a simple Gaussian function with a FWHM of 10 seconds and a peak
emission at 10 seconds \citep{Kouveliotou93}: 
\begin{equation} 
T_p(t) =
A_{pt} G[t,\mu = 10\;{\rm s},\sigma = 10 /(2 \sqrt{2 {\ln} 2})\;{\rm
s}]\;. 
\end{equation} 
The value of the constant $A_{pt}=1.00935\;{\rm s}^{-1}$ is determined
so that the Gaussian is normalized to 1 in the interval $0<t<\infty$.

We use the broken power law form given by \citet{Band_etal93} for the
frequency dependent component of the prompt emission:

$$F_E(E) = A_{pf} \left (\frac{E}{100 \mbox{keV}} \right )^{\alpha}
\mbox{exp} \left (-\frac {E}{E_0}\right) ,\;\;\;\;(\alpha - \beta)E_0
\ge E $$ $$F_E(E) = A_{pf} \left [ \frac { (\alpha - \beta)E_0}{100
\mbox{keV}} \right ] ^{\alpha - \beta} \mbox {exp} \left ( \beta -
\alpha \right ) \left ( \frac {E}{100 \mbox {ke V}} \right )^{\beta} ,$$
\begin{equation}
\;\;\;\;\;\;\;\;\;\;\;\;\;\;\;\;\;\;\;\;\;\;\;\;\;\;\;\;\;\;\;\;\;\;\;\;
\;\;\;\;\;\;\;\;\;\;\;\;\;\;\;\;\;\;\;\;\;\;(\alpha - \beta)E_0 \le E\;,
\;\;\;\;\;\;\label{PromptFlux} \end{equation}
where $E = h\nu$ is in units of keV, $(\alpha-\beta)E_0$ is the knee of
the power law, taken to be 300 keV, $\alpha$ and $\beta$ are the slopes
of the power law for $E < E_0$ and $ E > E_0$, respectively, and
$A_{pf}$ is a normalization constant. In our model, $\alpha = 0$ and
$\beta = -2$. Using these values makes the integral easy to solve
analytically:

$$ \int_0^{2E_0} \exp (-E/300 \keV) dE +$$ $$ \int_{2E_0}^\infty \left
(\frac {2E_0} {100} \right ) ^2 e^{-2} \left ( \frac E {100 \keV} \right
) ^ {-2} dE = 300(1+e^{-2}) \keV $$

yielding the normalization constant $A_{pf} = 2.935990\EE{-3}
\keV^{-1}$.

Thus the model for the prompt emission is \begin{equation}
\label{afterglowmodel} L_\nu (t,\nu,\theta) = E\sub{iso} F_E\left (\frac
{h\nu} {1000}\right) T(t) \left (\frac {\theta_0}{\theta} \right ) ^2
\end{equation} where $E\sub{iso}(\theta/\theta_0)^2$ is the total
isotropic energy output along the line of sight determined by $\theta$.
It is related to $L\sub{iso}(\theta)$ through $E\sub{iso}=\int_0^\infty
L\sub{iso}(t,\theta_0) dt$.

\subsubsection{Afterglow Model} The time component of the afterglow
emission is modeled as a smoothly broken power law \begin{equation}
T_{AG}(t) = \frac {A_{AGt}} { \left ( \frac{t}{t_b} \right ) ^ {-3} +
\left ( \frac {t}{t_b} \right ) ^{\beta} }\;, \end{equation}
with $\beta = 0.975$ and $A_{AGt}$ a fitting constant, picked so that at
peak emission $T_{AG} \left (t\sub{peak} \right ) = 1.$ The powerlaw
break is at $t_b = 100 $ s. We
analytically find the peak time to be $t\sub{peak} = 132.677$ s, and our
fitting constant was determined to be $A_{AGt} = 1.72796$.

To normalize the frequency dependent component we use the observation
that $R = 15$ for a z = 1 GRB. \citep{vanParadijs_etal00}. To fit this,
we used the R-Band Bessel filter zero point: $R=0$ at a flux density of
3080 Jy at $\nu\sub{R}=4.69\EE{14}$ Hz. This gives an $R=15 $ flux
density of .00308 Jy at the same frequency.  Using a luminosity distance
$d_L(z=1)=6626.5$ Mpc (for a cosmology with $\Omega_\Lambda = 0.73
,\Omega_M = 0.27$, and $H_0 = 71$ km/s/Mpc), we find a specific
luminosity $L_\nu(t\sub{peak}, \nu\sub{R}) = 1.62 \EE{32} $erg/s/Hz.
Since $T_{AG} (t\sub{peak}) = 1$, then $ L_\nu (t\sub{peak} ,\nu\sub{R})
= F_{\nu}(\nu\sub{R}) = A_{AGf} \nu\sub{R}^{-\alpha} = 1.62 \EE {32}
\erg / \s /\Hz$ with the typical value of $\alpha = 0.65$ 
(corresponding to an index $p=2$ of the electron distribution). This
readily yields $A_{AGf} = 5.57\EE{41}$. Thus, the spectral component is
\begin{equation} 
F_{AG} (\nu) =  5.57\EE{41} \nu^{-0.65}\; {\rm erg\;s^{-1}\,Hz^{-1}}\;, 
\end{equation} 
giving the following model for the afterglow component
\begin{equation} L_{AG} (\nu,t,\theta) = F_{AG} (\nu)T_{AG} (t) \left
(\frac { \theta_0} {\theta} \right ) ^2 \end{equation} with
$F_{AG}(\nu)$ and $T_{AG}(t)$ defined as we just discussed. This
corresponds to a total effective isotropic energy output across the range  1
eV to 10 keV of about $10^{52} $ ergs, about $10\%$ of the prompt
emission.

\subsection{ Computation of the Transmittance} \label{transcomputation}
During the radiative transfer calculation, we keep track of the
time-dependent density of each ion present in each radial bin, hence obtaining
the evolving ionization profile as a function of radius for each ion of
interest. We track the following ions: HI, HeI, HeII, CII, CIII, CIV,
SiII, SiIV, MgI, MgII, FeII, and OII.  This, combined with our
radial velocity and temperature profile, is sufficient to generate a
time-dependent transmittance, which we compute at infinite instrument
resolution.

To calculate the transmittance, we compute the Voigt line profile
function for each individual line, \begin{equation} \phi \left (\nu
\right ) = \frac{H(a,u)}{\Delta \nu_D \sqrt{\pi}}  , \label{voigt}
\end{equation} where $H(a,u)$ is the Voigt function: \begin{equation}
H(a,u) = \frac{a}{\pi} \int_{-\infty}^{\infty} \frac {e^{-y^2} dy}{a^2 +
(u-y)^2} \end{equation} with $ a = {\Gamma}/{4 \pi \Delta \nu_D}$ and $
u \equiv {(\nu - \nu_0)}/{\Delta \nu_D}$.  Here $\Delta \nu_D$ is the
Doppler width and $\Gamma$ is the transition rate, which is approximated
by $\gamma$, the spontaneous decay rate \citep{RybickiLightman79}.  
The cross section for a given line is \begin{equation}
\sigma_\nu = \frac{\pi e^2}{mc} f_{12} \phi (\nu), \end{equation}
from which we can calculate the optical depth $\tau(\nu)$. The
transmittance is thus \begin{equation} Tr(\nu) = Tr_0(\nu)*e^{-\tau} ,
\end{equation} where $Tr_0$ is its value  before
being affected by the opacity of the given line.  

The transmittance was calculated at each radial bin taking into
account 21 resonant transitions from the above mentioned ions; thanks
to the inclusion of the Doppler effect, any structure present in the
velocity profiles is correspondingly reflected in the simulated
spectra. We limited ourselves to 21 resonant transitions since this
was the most computationally expensive part of our study. The lines
were selected based on observed GRB spectra \citep{Fiore_etal05} rest
wavelength, and oscillator strength $f_{12}$. The lines present
in our calculations are listed in Table \ref{LinesChosen}. The
reported $f_{12}$ values were calculated from the Einstein A
coefficients given by \citet{NIST}.  It is worth noting that we
only treat lines that arise from resonant transitions in the ground
state of each ion, although fine-line transitions are commonly observed in
GRB afterglow spectra (Vreeswick et al. 2003, 2007; Prochanska et al. 2006; 
D'Elia et al. 2009a,b). 
At present time, there is no time-dependent,
radiative transfer code that treats self-consistently transitions from
both the ground and the excited levels, as well as the time-dependent
evolution of the dust grains. However, we estimated the effect of our
approximation by using a time-dependent code (van Adelsberg et al. in
prep.) which computes the evolution of the excited states of the ions
under the influence of a time-variable radiation field.  By running
this code with the luminosity function in \S2.3 and viewing angle
$\theta=0^\circ$, we found that, for ions such as SiII and FeII, in
the innermost region of the wind ($\sim 1$ pc) the ground level
depopulates to about 20\% on a timescale of several seconds, while in
the outermost regions of the wind ($\sim 10$ pc), it depopulates to
about 30\% on a timescale of several hundred seconds. The depopulation
timescales become longer at larger viewing angles. Therefore, given
the timescales over which the transient high-velocity features live
(\S 3), the correction to the strength of the ground-level resonant
lines due to the depopulation of the ground level is not expected to
be significant.  In the case of CIV, we found the population of the
excited levels to be negligible across the full extent of the
wind.

\begin{table*}
  \caption{Lines and Oscillator Strengths of the 21 Transitions
considered for the calculation of our transmittance spectra. }
\centering
\begin{tabular}{ c c c c c c} \hline\hline Ion & $\lambda_0[\AA]$ &
$f_{12}$ & Ion & $\lambda_0[\AA]$ & $f_{12}$\\
\hline
HI & 1215.67 & 0.278 &
CII & 903.962 & 0.336 \\
CII & 1334.53 & 0.127 &
CIII &977.03& 0.759\\
CIV & 1548.20 & .190&
CIV& 1550.77 & 0.0952 \\
MgI & 2852.50 & 1.80 &
MgII & 2796.35 & 0.566 \\
MgII &2803.53 & 0.303 &
SiII & 1190.42 & 0.139 \\
SiII & 1193.29 & 0.574 &
SiII & 1260.42 & 1.224 \\
SiII &1304.37& 0.0928 &
SiII & 1808.01 & 0.00248 \\
SiIII & 1206.50 & 1.67 &
SiIV &1393.76 & 0.512 \\
FeII & 2344.21 &  0.11 &
FeII & 2374.46 & 0.028 \\
FeII & 2382.77 & 0.38 &
FeII & 2586.65 & 0.065\\
FeII & 2600.17 & 0.22 & & & \\
\end{tabular} \label{LinesChosen} \end{table*}

\section{Results}

Our calculations were performed both for a non-dusty CSM and for a dusty
CSM. In the non-dusty case,  most of the inner region is completely or
near-completely ionized before the burst occurs due to the ionizing
photons from the progenitor star, vastly reducing the signatures of the
high-velocity wind close to the star. These features, that is
high-velocity lines from low-ionization elements, are instead visible in
the case in which metals are initially locked into dust grains and later
recycled into the ISM by the GRB radiation. The absorption features from
the outer shell, on the other hand, are very similar for the non-dusty
and the dusty scenarios. Therefore, we present in the following detailed
results only for the case of a GRB exploding in the dusty CSM of a
massive star.

\begin{figure*} 
\begin{center} \includegraphics[scale =
.95,trim=3cm 0 -3cm 0, clip ]{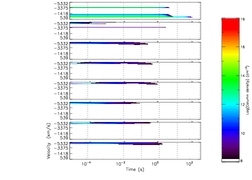}
\hspace{-3.0 cm} \caption{Observable column density at different
velocities versus time at $\theta = 0^ \circ$. Each panel
corresponds to a different ion. They are, in order from top to bottom,
HI, HeII, SiII, SiIII, SiIV, CII, CIV, FeII.   The color/shade
indicates the observable column density. The y axis indicates at what
velocity the lines will be observed. Each horizontal line element
corresponds to a velocity interval of 54 km/s.  The x axis shows time. 
The vertical dotted white lines indicate where we show sample  spectra. 
} \label{IonDensity_1} \end{center} \end{figure*}

\begin{figure*} 
\begin{center} \includegraphics[scale =
.95,trim=3cm 0 0 0, clip]{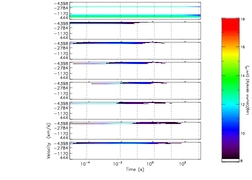}
\caption{Similar to figure \ref{IonDensity_1} but for $\theta = 25^
\circ$, and with each horizontal line element corresponding to a
velocity interval of 44 km/s. Again, the different ions are, from top to
bottom, HI, HeII, SiII, SiIII, SiIV, CII, CIV, FeII. }
\label{IonDensity_3} \end{center} \end{figure*}

\begin{figure*} 
\begin{center} \includegraphics[scale =
.95,trim=3cm 0 0 0, clip]{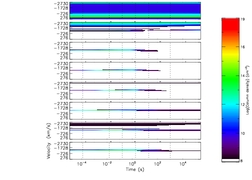}
\caption{Similar to figure \ref{IonDensity_1} but for $\theta = 80^
\circ$, and with each horizontal line element corresponding to a
velocity interval of 28 km/s. The different ions are, from top to
bottom, HI, HeII, SiII, SiIII, SiIV, CII, CIV, FeII.}
\label{IonDensity_6} \end{center} \end{figure*}

 Figures \ref{IonDensity_1}, \ref{IonDensity_3}, and
\ref{IonDensity_6} display the results of the time dependent
ionization profiles for HI, HeII, SiII, SiIII, SiIV, CII, CIV, and FeII,
along lines of sights corresponding to $\theta =0^\circ$, $25^\circ$, and
$80\degrees$. To incorporate velocity information, we show a shaded
column density map for each line of sight and a few key ions, with
color/shade indicating the observable column density along the line of
sight, the y-axis indicating the velocity, and the x- axis indicating
time. A negative velocity here represents a blue-shifted
velocity towards the observer.  Note that the maximum velocity does vary
with angle, as a result of the anisotropic nature of the CSM produced by
a rotating star.  The displayed times range logarithmically from about
$1\EE{-5}$ seconds to $2.5\EE5$ seconds.

There are some common features that are interesting to note.  In
particular, the column densities of the gaseous components of the
low-ionization states of C, Fe and Si, which are locked into dust grains
before the burst goes off, are negligible at $t=0$ (explosion time),
they then increase as the grains are destroyed by the burst radiation
and recycled into the ISM, and eventually decrease as the ions are
photoionized.  This time evolution sequence is much faster at small
viewing angles, given the stronger radiation field from the GRB.

Figures \ref{SpectrumR1_a} through \ref{SpectrumR3_a}  show
several snapshots (between $t\sim 10^{-4}$ s and $t\sim 10^3$~s) of
the normalized transmission spectra for each of the three representative
lines of sight.  The times in these figures correspond to the vertical
dotted lines in the aforementioned column density plots. We do not
include information at later times because the spectra have stopped
evolving at this point, as is apparent from Figures {
\ref{IonDensity_1}, \ref{IonDensity_3}, and \ref{IonDensity_6}.}

\begin{figure*} 
\begin{center} \includegraphics[scale =
.75]{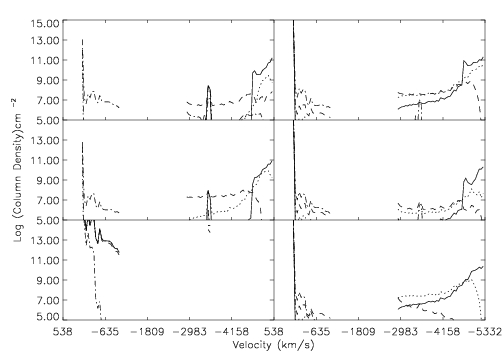} \caption{Number Density [$\log $ cm$^{-2}$]
versus velocity [km/s]. Each panel represents a cross-section of Figure
\ref{IonDensity_1} at the times t = 0.17 s (solid), 1.80 s (dots), 20 s
(dashes), and 21000 s  (dashes-dots). The ions presented are  HI (lower
left), Si II (lower right), SiIV (middle left), CII (middle right),
CIV(upper left), and FeII (upper right). } \label{Cross_1} \end{center}
\end{figure*}

\begin{figure*} 
\begin{center} \includegraphics[scale =
.75]{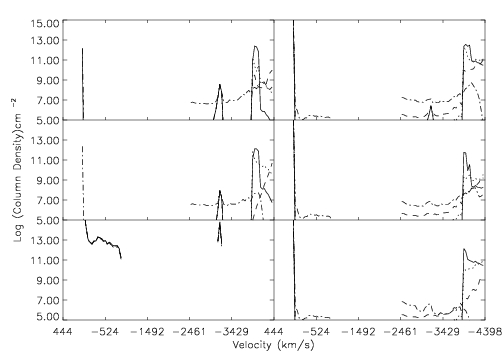} \caption{Cross sections of Figure
\ref{IonDensity_3} of the same ions and at the same times as in Fig.
\ref{Cross_1}.} \label{Cross_3} \end{center} \end{figure*}

\begin{figure*} 
\begin{center} \includegraphics[scale =
.5]{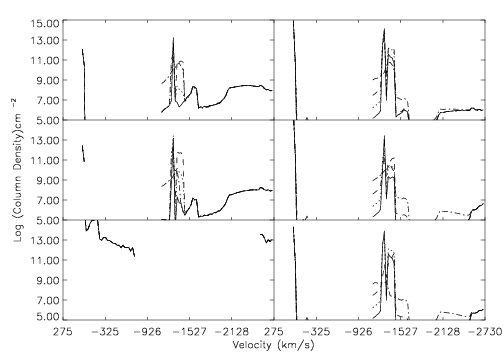} \caption{ Cross sections of Figure
\ref{IonDensity_6} of the same ions and at the same times as in Fig.
\ref{Cross_1}.  } \label{Cross_6} \end{center} \end{figure*}

Additionally, we include cross sections across the column density plots
for the time steps shown in the sample spectra (Figures \ref{Cross_1},
\ref{Cross_3},and \ref{Cross_6}). These simply show, perhaps in a
simpler manner than the column density maps, the column densities vs.
the velocity of the gas for a few selected ions.
We indicate the most notable features in section \ref{manifestation}.

\subsection{Manifestation of the Velocity Structure. }
\label{manifestation}

To keep these results in context, it is helpful to refer to Figs.
\ref{InitProfile} and \ref{Velocity2D}, since the velocity structure
depends on $\theta$, and also it is helpful to note the location of
certain velocity structures since this has a direct impact on the
ionization timescale. For example, along all lines of sight, the free
streaming region extends from our minimum radius of $3\EE{16}$ cm, and
experiences a slow variation in velocity from there to a distance of
about $5\EE{18}$ cm.  This is the innermost area of the CSM in our
simulation. The inner subset of this region, up to about $1\EE{17} $ cm,
has the highest number density, and a temperature around $5\EE{5}$ K
(Figure \ref{InitProfile}). This is the region where  the high-velocity
features seen in the  spectra are produced. For example, for $\theta =
0^\circ$ (Figure \ref{IonDensity_1}), this accounts for the 3500 km/s
feature for HI, as well as the $\approx$ 5000 km/s component present in
several of the other ions.

As Figure \ref{InitProfile} indicates, there are indeed isolated
structures at specific velocities. The most dominant one is perhaps the
least interesting, the outer shell which is nearly at rest ($v_r \la25$
km/s), and is expected by the most basic stellar wind models.  This is
observable as the high density feature at the bottom of each shaded
column density plot.  Equally ubiquitous is a large bubble of shocked
wind, which is visible in Figure \ref{InitProfile} as the sudden drop in
velocity at about $r = 1.5\EE{19}$ cm. This region occupies the majority
of our simulation space, and because of its very hot, $\sim
1\EE{8}-1\EE{9}$ K temperatures, none of the ions that we track are
present in this region.


We present the spectra for four different ions where we seek to
determine if the velocity structure becomes visible. First, we note that
Hydrogen is highly ionized in the inner region and thus is a poor tracer
of velocity structure. However, the outer shell does contribute a large
hydrogen component and the near 1215.67 $\AA$ rest frame Ly-$\alpha$
feature can thus act as a diagnostic of the outer shell.

\begin{figure*} 
\begin{center} \includegraphics[scale =
.95]{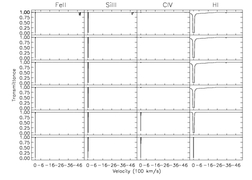} \caption{ 
Transmittance versus velocity (100 km/s) for Fe II  2382.04 $\AA$
(left), SiIII 1206.50 $\AA$ (left-middle), CIV 1550.77 $\AA$
(right-middle) and HI 1215.67 $\AA$ (right) at  $\theta = 0^ \circ$ and
at times $ t = 10^{-4}, 0.17,1.80, 20$,  and $1040$ seconds (from top to
bottom). Each spectrum is normalized and is scaled identically.   }
\label{SpectrumR1_a} \end{center} \end{figure*}

\begin{figure*} 
\begin{center} \includegraphics[scale =
.95]{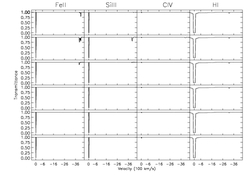} \caption{ 
Similar to Figure \ref{SpectrumR1_a} but for $\Theta = 25^\circ$. 
}
\label{SpectrumR2_a} \end{center} \end{figure*}

In the $\theta = 0$ case, (Fig. \ref{SpectrumR1_a}), we see some
higher velocity components for ions associated with dust such as SiIII,
FeII, and CIV. These appear only at early times, when the burst
initially destroys the dust, releasing Si and Fe. However, at this
head-on angle, the intensity of ionizing radiation is high, quickly
ionizing the released ions and destroying this signature of the high
velocity wind almost immediately.

\begin{figure*} 
\begin{center} \includegraphics[scale =
.95]{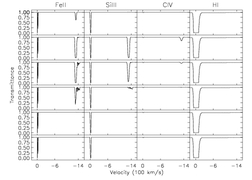} \caption{ 
Similar to Figure \ref{SpectrumR2_a} but for $\Theta = 80^\circ$.}
\label{SpectrumR3_a} \end{center} \end{figure*}

In the $\theta = 25 \degrees$ line of sight, the rest frame
Ly-$\alpha$ line exhibits a behavior similar to the $\theta = 0^\circ$
case, but on longer timescales. Rest frame and near rest frame
features also exhibit a very similar behavior as in the $\theta =
0^\circ$ case.  Of particular note is Figure {\ref{SpectrumR3_a}},
corresponding to the $\theta=80^\circ$ viewing angle, in which we see
finely structured FeII and SiIII lines due to the release of gas from
dust in the high velocity region of the CSM. These survive on the
order of a few seconds before being ionized to higher states.

\section{Summary and Discussion}

In this paper we have presented a comprehensive calculation of the
spectra of a typical GRB which explodes in the CSM created by a massive,
chemically homogeneous, low-metallicity rotating star. We have explored
both a dusty CSM as well as a dust-free one. These stars are strong
candidates for the typical progenitors of long GRBs; therefore, the
models presented here, in combination with an average GRB luminosity and
spectrum, are expected to yield the typical expectations for the spectra
during both the prompt and the afterglow phase of the GRB. We
emphasize, however, that an accurate prediction can be obtained only from a
tailored time-dependent simulation for each particular burst, and hence our
results should rather be seen as an example of the overall evolution, whose
details will then depend on the spectrum and luminosity of the specific burst.

We have performed the calculation of the transmittance for different
lines of sights with respect to the GRB jet axis (assumed to coincide
with the rotation axis of the star), thus exploring the latitudinal
variation of the spectra, which include the $\theta$-dependence of both
the CSM profiles as well as of the GRB luminosity.

The highest velocity components of the CSM, reaching velocities above
5000 km/s, exist predominantly at small viewing angles (with respect
to the rotation axis of the progenitor). This is where the burst is
most luminous, and thus ionization timescales are very fast. We do see
the presence of several ions at these velocities, but this region
becomes completely ionized in a fraction of a second, and an
actual observation of the high velocity features seems hopeless, at
least for the star considered in this study.

Lines of sight at higher latitude are characterized by smaller maximum
velocities, with the maximum being around 2800 km/s in the case of a
line of sight at $80^\circ$. {In our model, even at these latitudes,
ions} with lines around 1000 km/s cannot survive the entire burst
without being photoionized.  From a purely theoretical prospective,
the time variability of the high velocity lines, and in particular, the
transition from a nearly featureless spectrum to high column density
lines back to a featureless or weak lined system like the one observed in
our study, is a clear signature that the lines are associated with the
CSM, and furthermore, that they are due to the destruction of dust in
the CSM created by the progenitor star. Practically, however,
observational detection of such variability with current instruments is
not feasible.

Observationally, high-velocity lines have been reported in GRB~021004 by
\citet{Fiore_etal05} after about 16 hr from the trigger.  From the
analysis presented in this paper, we find that these features, if
associated with the free-streaming wind of a progenitor star like
the one discussed here, could have survived till these late times only
if the burst had a very low luminosity. Since this is
not supported by observations (the luminosity of GRB~021004 was rather
high), the only way to associate these lines to a wind is by requiring a
very special event, in which the progenitor star would possess an
unusually strong wind, and it were to explode in a very low density
medium. Alternatively, the observed late-time, high-velocity lines could
be simply due to intervening absorbers (Chen et al. 2007).

{In summary, during} the initial moments of the GRB explosion, the
optical absorption spectra present a very rich phenomenology due to
the different speeds at which the wind components move away from the
progenitor side \citep{vanMarle_etal08}. If we were able to observe
these early phases, we could derive a wealth of information on
the properties of the progenitor stars of GRBs.  More generally, this
would be of great importance to the study of the final stages of
stellar evolution.  Observationally, however, it is extremely
challenging to catch variability on such small timescales, shorter
even than the time it usually takes to localize the burst. 
Therefore we conclude that, if the massive stars progenitors of
GRBs are indeed of the kind studied here (as suggested by theoretical
investigations), then the high-velocity lines produced in their winds are
practically invisible, even for very low-luminosity bursts (such as the
$\theta=80^\circ$ case that we considered). The only observational
signatures of the GRB progenitor stars are imprinted in non-variable
absorption lines produced in the termination shock of the wind. 
Previous detections of high-velocity lines must therefore be 
due to intervening absorbers, unless the GRB progenitor star had extreme
properties well outside of the range considered here.

\section{acknowledgments} We thank Sung-Chul Yoon and Norbert Langer
for the use of their stellar evolution model, and Matt van
Adelsberg for his calculations of the fraction of excited levels for
the most important ions that we considered. 
We also thank a very careful and thoughtful referee, whose
comments greatly helped the presentation of our manuscript.

\end{document}